\documentclass[twocolumn,aps,showpacs]{revtex4}
\usepackage[dvips]{graphicx}
\usepackage{amssymb}
\usepackage{bm}
\begin{document}
\title{
Oceanic coastline and super-universality of percolation clusters.
}

\author{Jaan Kalda}
\address{\dag\ 
%\affiliation{
Institute of Cybernetics,
Tallinn Technical University,
Akadeemia tee 21,
12618 Tallinn,
Estonia
}

\begin{abstract} 
New fractal subset of a rough surface, the ``oceanic coastline'', is defined.
For random Gaussian surfaces with negative Hurst exponent $H<0$, 
``oceanic coastlines'' are mapped to the 
%According to the mapping between the lattice percolation and continuous  
percolation clusters of the (correlated) percolation problem.
In the case of rough self-affine surfaces ($H\ge 0$), the fractal dimension of the ``oceanic coastline'' $d_c$ is 
calculated numerically as a function of the roughness exponent $H$ (using a novel technique 
of minimizing finite-size effects). For $H=0$, the result $d_c \approx 1.896$ 
coincides with the analytic value for the percolation problem (91/48), suggesting a super-universality
of $d_c$ for correlated percolation problem.
\end{abstract} 

\maketitle

Random surfaces are common objects which can be found in various situations;
good examples are
deposited metal films \cite{Kondev2},
fractured surfaces \cite{Sahimi,Bouchaud}, 
streamfunctions of two-dimensional (2D) turbulent flows \cite{MBI} etc.
For various applications, it is important to know the statistical geometrical 
properties (statistical topography) of such surfaces. This problem has been studied 
with a considerable success, both numerically and analytically \cite{Kondev2,Sahimi,Bouchaud,MBI,IK,Kondev1,Olami}. 

Here we introduce a new scale-invariant measure of such surfaces, the fractal dimension of the ``oceanic coastlines''.
An ``oceanic coastline'' includes the outer perimeter of the ocean (the ``continental coastline''), and
the coastline of ``oceanic islands'', i.e.\ of such islands, which are adjacent to the largest (possibly infinite) 
connected flooded region, when water floods the surface up to a certain level; ``oceanic islands'' are to be 
opposed to such islands, which are inside the lakes. 
Thus, each flooding level defines its own ``oceanic coastline''.
Here we show that for the so called ``monoscale'' 
random surfaces (with bounded height, cf.\ \cite{IK}), the ``oceanic coastlines'' can be mapped to the percolation clusters of the 
correlated percolation problem. For rough surfaces (with unbounded height), the fractal dimension $d_c$ is calculated 
numerically, using a novel technique of minimizing the finite-size effects. The results suggest that 
for the correlated percolation (and hence for the ``monoscale'' surfaces), there is a super-universality, $d_c\equiv 91/48$.

An important characteristic of random surfaces [given by the surface height $\psi(x,y) \equiv \psi(\bm{r})$ 
over the two-dimensional plane] is the Hurst exponent $H$; in the case of 
Gaussian surfaces, all the statistical topography scaling exponents are functions of $H$.
The most intuitive definition of the Hurst exponent is given by the scaling law of the surface height drop
at a distance $a\equiv |\bm{a}|$ (c.f.~\cite{Mandelbrot}):
\begin{equation}
	\left< (\psi(\bm{r})-\psi(\bm{r}+\bm{a}) )^2\right>  \propto |\bm{a}|^{2H}.
\end{equation}
Here angular braces denote averaging over an ensemble (or over the radius vector $\bm{r}$); 
we assume that $a$ is larger than the lower cut-off scale $a_0$, and $0\le H\le 1$.
More general definition (not bounded to $0\le H\le 1$) of the exponent $H$ is given via the power spectrum $P_{\bm{k}}$ 
[with $\left< \psi_{\bm{k}} \psi_{\bm{k}^\prime}\right> = P_{\bm{k}}\delta(\bm{k}+\bm{k}^\prime)$, where $\psi_{\bm{k}}$ is the
Fourier transform of the surface height; c.f.~\cite{MBI}]:
\begin{equation}
P_{\bm{k}} \propto |\bm{k}|^{-2H-2}\;\;\; \mbox{for}\;\;\;  |\bm{k}| \ll a_0^{-1}.
\end{equation}

According to the value of $H$, surfaces can be divided into three classes \cite{IK}: \\
{\em (i)} $H<0$: ``monoscale'', surface looks like a landscape with hills of approximately same size, the surface
height is bounded;\\
{\em (ii)} $0\le H < 1$: ``rough'', looks like rocky mountains with smaller and higher peaks and unbounded height;\\
{\em (iii)} $H \ge 1$: ``smooth'', the largest scale defines the slope of the surface.

For ``monoscale'' surfaces, there is a simple mapping 
between statistical topography and percolation problem,
developed by Ziman and Weinrib \cite{Ziman,Weinrib1}
[the minima of the surface define the sites of an 
irregular percolation lattice; the saddle-points $S_{ij}$ (between $i$-th and
$j$-th minima)  correspond to the bonds of the lattice (which can be drawn as 
the steepest descend paths from $S_{ij}$ to the  $i$-th and
$j$-th minima); when the surface is flooded 
by water, a bond is declared broken if the corresponding saddle is (``dry''), i.e.\ above the water level;
the water level itself is the counterpart of the bond-keeping probability $p$].
This mapping can be also applied to the case of rough surfaces. Then, however, the lattice problem is no more the
(correlated) percolation problem. In particular, there is no critical probability $p=p_c$ at which  the 
global connectivity disappears. Indeed if we ``look far enough'', we can find as 
deep (hence also as wide) connected wet region as we want, because the height is unbounded.
Therefore, the mapping between the percolation problem and statistical 
topography allows us to extend various concepts of the percolation problem outside the range of applicability of the 
very percolation problem. 

A straightforward example here is the hull of a percolation cluster: this is defined as the outer perimeter of a
percolation cluster. Since the percolation cluster corresponds to a connected wet region, its outer perimeter corresponds
to an isoline of the surface. It has been shown that for uncorrelated percolation, its fractal dimension $d_h=7/4$ \cite{Saleur}.
Note that percolation with weak correlation (with $H<-3/4$) belongs to the same universality class as the uncorrelated
percolation \cite{Weinrib2}.
Meanwhile, for rough surfaces, $d_h\approx 1.5-0.5H$ \cite{JK}, with an exact result $d_h=1.5$ for $H=0$ \cite{Kondev1}.
Therefore, for $-3/4<H<0$, one can expect a non-trivial decreasing function $d_h(H)$, with $d_h(-3/4)=1.75$ and $d_h(0)=1.5$.
Further, for the percolation problem, one can distinguish between the hull and the unscreened perimeter. This is because
the hull is very twisted, and typically, there are narrow ``necks'' of peninsulas separating ``gulfs''. We can postulate that these 
``necks'' are too narrow, so that the coastline of the respective peninsulas is ``screened'';
the rest of the coastline is called  ``unscreened perimeter''.
For percolation clusters, the fractal dimension of 
the unscreened perimeter is $d_u = 4/3$ \cite{Saleur}. Since for rough surfaces, narrow ``necks'' are untypical, $d_u(H)=d_h(H)$ 
for $H\ge 0$. Comparing $d_u(-3/4)=4/3$ and $d_u(0)=1.5$, one can expect an increasing function $d_u(H)$ for 
$-3/4<H<0$.

Now we are ready to turn to the percolation cluster. Each bond of the cluster is touched by the perimeter (hull) from both sides;
thus, the length of the net perimeter scales as the cluster size.
However, the perimeter touching it can be either external  or internal, see Fig.~1. The sum of external and internal 
perimeters is the  sum of {\em (a)} the coastline between the ``ocean'' (=cluster) and  the ``continent'' (=infinite cluster of 
broken bonds), and {\em (b)} the coastline between 
the ``ocean'' and the ``islands'' (=broken bond clusters inside the cluster of undamaged bonds); we call this the ``oceanic coastline''.
Note that the coastlines of the ``lakes'' inside the ``islands'', ``islands'' inside the ``lakes'' inside the ``islands'', etc.,
do not touch the ``ocean'', i.e.\ the bonds of the cluster, and are not relevant here (including these coastlines, too, 
would result in the set of all the isolines at the given height, the fractal dimension of which 
$d_f=2-H$ for $0\le H \le 1$; for $H\le 0$, there is a super-universality, $d_f \equiv 2$ ).
\begin {figure}[!t]
\includegraphics{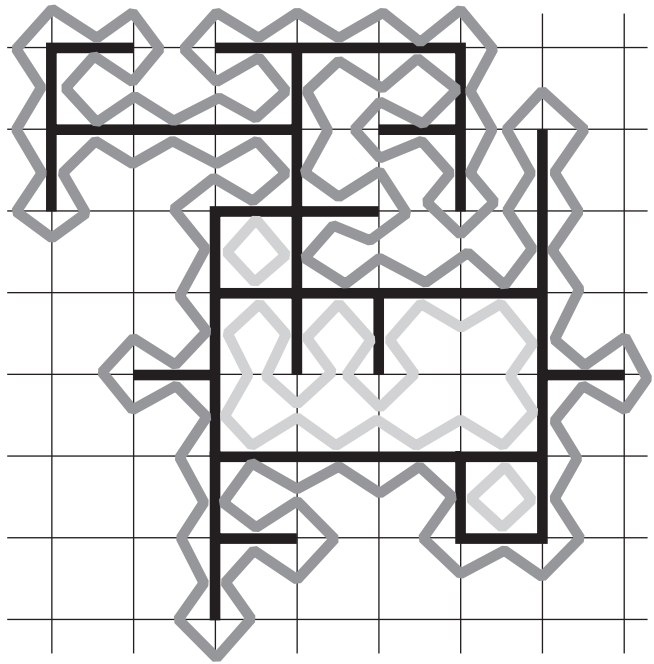}
\caption{All the bonds of a percolation cluster (black bold lines) are touched from both sides, either by external 
perimeter (hull, dark grey lines) or internal perimeter (light grey lines).} 
\end {figure}

In order to calculate numerically the fractal dimension of the ``oceanic coastlines'', 
we have opted for the following procedure.
First, 1+1-dimensional [(1+1)D] random surfaces 
(which are assumed to belong to the same universality class as the statistically 
isotropic 2D surfaces) were generated on the lattice of the four-vertex model \cite{JK}. The 
one-dimensional discretized fractional Brownian functions $f_H(i)$ were generated as fractional integrals of uncorrelated random
sequence of ``spins'' $s_i=\pm 1$ as follows. The aim function $g_H(i)$ is found as 
\begin{equation}
	g_H(i) = \sum_{j=0}^j s_j |i-j|^{H-0.5}.
\end{equation}
Further, the discretized function $f_H(i)$ is found incrementally, starting from $H=0$, by rounding $g_H(i)$ to such a nearest integer that
\begin{equation}
	f_H(i)-f_H(i-1) = \pm 1.
\end{equation}
Finally, the (1+1)D function is given by the superposition of two 1D-functions,
\begin{equation}
F_H(i,j)= f_{H1}(i)-f_{H2}(j).
\end{equation}

After generating a 2D surface on a square polygon of side length $N$, we started ``flooding'' the polygon. To this end, 
at the perimeter of the polygon, the lowest point $P_0$ 
was found. Further, ``water'' was incrementally ``injected'' into that point, until the flooded 
region connected opposite edges of the polygon (either left and right or top and bottom), see Fig.~2. 
Such a procedure guarantees that only the ``ocean'' (and not the ``lakes'') becomes wet.
At that stage, three quantities were 
recorded: $L_1$ --- the coastline length, $L_2$ --- the number of cells touching the coastline, and $L_3$ --- the coastline length 
immediately before achieving the critical flood level. The procedure was repeated (for each polygon size and Hurst exponent value) 
$10^8$ or more times. The covered polygon sizes are given by formulae $N=128\cdot 2^k$ and N=$196\cdot 2^k$, with $k=1,2,3,4$, and $N=4096$ (the largest 
polygon size was used only for $H=0$). The simulations were performed on the cluster of ten AthlonXP-1700 workstations during three months, using 
an assembly-optimized code.
\begin {figure}[!b]
\includegraphics{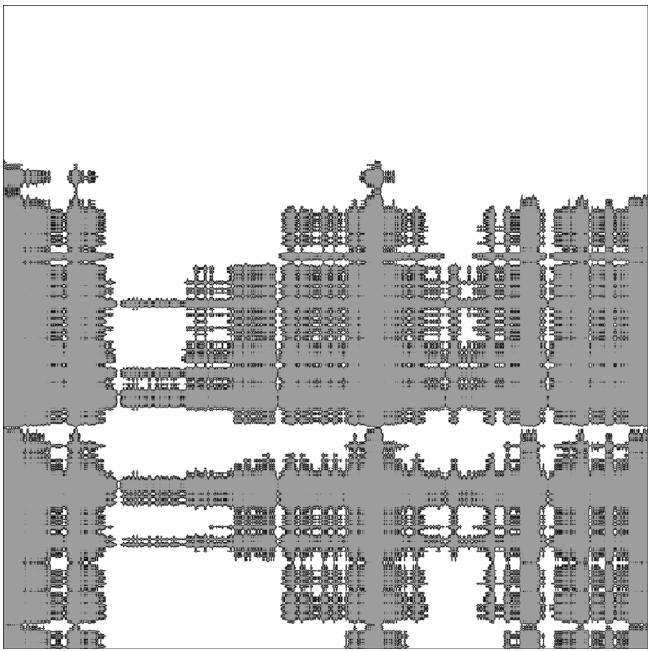}
\caption{The polygon ($N=2048$, $H=0.5$) has been flooded so that the flooded region (grey area) connects left and right edges of the
polygon. The striped patterns (in horizontal and vertical directions) are due to the (1+1)D geometry of the surface.}
\end {figure}

\begin {figure}[!t]
\includegraphics{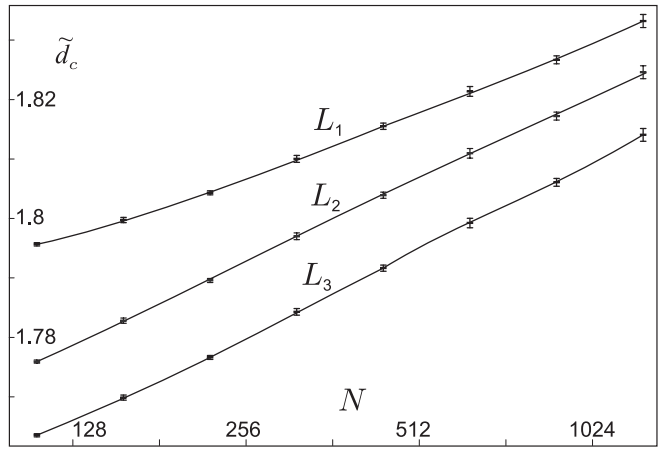}
\caption{Finite-size effects for $H=0$ give rise to a very slow convergence of the estimated fractal dimension 
$\tilde d_c(N) = \ln [l_i(N_{k+1}/l_i(N_k)] / \ln(N_{k+1}/N_k)$, $N=\sqrt{N_kN_{k+1}}$.}
\end {figure}
It was expected that the finite-size scaling of the average values $l_i(N) = \left<L_i(N)\right>$
allows us to estimate the fractal dimension $d_c$. Such an asymptotic behavior was observed, indeed,  for larger Hurst 
exponents ($H \ge 0.375$). However, for smaller values of $H$ (and in particular, for $H=0$), the convergence 
$l_i(N) \to AN^{d_c}$ was very slow, see Fig.\ 3 (here $A$ denotes a constant).
Consequently, if we assume that $l_i$ can be expanded asymptotically,
\begin{equation}
	l_\nu=\sum_{\nu=0}^\infty A_{\nu i} N^{\alpha_{\nu i}}, \alpha_{(\nu+1)i} < \alpha_{\nu i} , 
\end{equation}
the non-leading exponents ($\nu \ne 0$) have to be very close to the leading term. In order to resolve this difficulty,
we made an assumption that for the three recorded quantities $l_i$ ($i=1,2,3$), the three leading 
exponents ($\nu=0,1,2$) are the same, $\alpha_{\nu i} = \alpha_{\mu i}$. The validity of this assumption is evident 
in the case of the leading exponent $\alpha_{\nu 0} \equiv d_c$, but not so clear for the other exponents.
However, if we accept the assumption, it can be verified later using the numerical results.
Thus, if we assume that $\alpha_{\nu i} = \alpha_{\mu i}$, one can construct such a linear 
combination of the three quantities $l_i$, for which two terms (out of the three leading ones) cancel
out. It suffices to find a root-mean-square fit by minimizing the expression
 \begin{equation}
 \begin{array}{l}
 S(d)=  \\
 \min_{A_1,A_2, A_3}\sum_{i=i_0}^{i_{\max}} \left[\frac{N^d(i)-A_1l_1(i)-A_2l_2(i)-A_3l_3(i)}{\sigma_i}\right]^2.
 \end{array}
%\begin{equation}
%S(d)=  \min_{A_1,A_2, A_3}\sum_{i=i_0}^{i_{\max}} \left[\frac{N^d(i)-A_1L_1(i)-A_2L_2(i)-A_3L_3(i)}{\sigma_i}\right]^2.
\end{equation}
Here, $\sigma_i$ is the variance of the expression $A_1l_1(i)-A_2l_2(i)-A_3l_3(i)$, and can be expressed via the individual variances 
of the quantities $l(i)$, and the pair-wise correlation coefficients; all this is easily calculated, using the
Monte-Carlo simulation results.
The resulting curves $\log_{10} [S(d)/(n-4)]$ are given in Fig.~4; 
here $n$ is the number of terms in the sum $S(d)$ (so that $n-4$ is 
the number of degrees of freedom). Note that the two smallest data-points (with $N_0=128$ and $N_1=196$) have been excluded from
the sum, otherwise, the residual squares would have been unacceptably large [resulting in $S/(n-4) \gtrsim 1$].
Evidently, for $N \le 196$, the higher ($\nu \ge 3$) terms of the expansion (6) become important.
The existence of three sharp minima for all the curves in Fig.~4 confirms the assumption which was made about the equality of the leading exponents.
Indeed, if the second and third asymptotic expansion terms of $l_1$, $l_2$, and $l_3$  were linearly independent,
they would not cancel out in Eq.~(7), and there would be no minima with $S/(n-4) \sim 1$.

\begin {figure}[!b]
\includegraphics{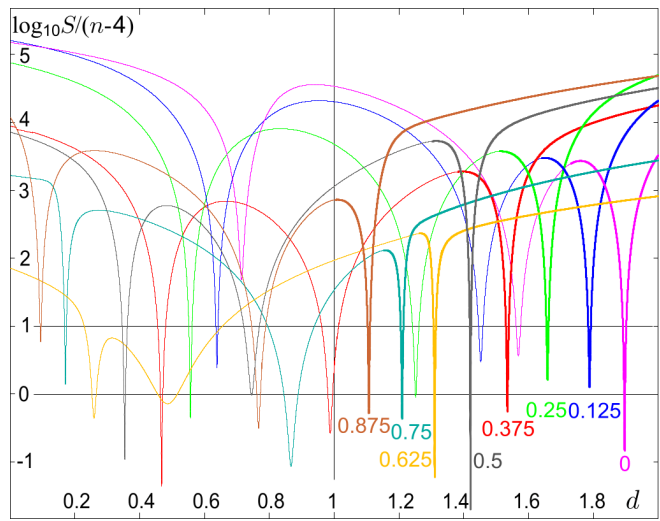}
\caption{The quantity $\log_{10} [S(d)/(n-4)]$ [see Eq.\ (7)] is plotted versus $d$ for different 
values of the Hurst exponent $H$ (each curve is labeled with a number indicating the value of $H$). 
For all the curves, three minima are clearly distinct. At these minima, $d=\alpha_i$, where
$\alpha_i$ is the $i$-th leading exponent of the asymptotic expansion of the quantities $l_1$, $l_2$, and $l_3$.}
\end {figure}

This novel method of interpreting the finite-size simulation results allowed us to calculate not only the fractal dimension of 
the ``oceanic coastlines'' (see Fig.~5), but also the second and third exponents in the asymptotic expansion of the length of 
``oceanic coastlines'' (Fig.~6). 
The values $d_c = 1.8965 \pm 0.0025$ and $\alpha_1 = 1.58 \pm 0.01$ for $H=0$ are relatively close to each other, exactly as it was 
expected, judging by the finite-size scaling  in Fig.~3.
\begin {figure}[!b]
\includegraphics{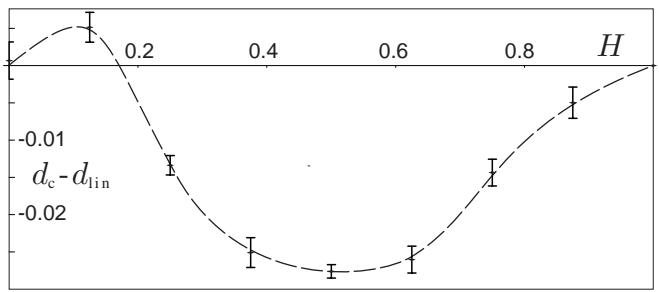}
\caption{The mismatch between a linear estimate $d_{lin}=\frac{91}{48}-H\frac{43}{48}$ and the fractal dimension of 
the ``oceanic coastline'' $d_c$ is plotted versus the Hurst exponent $H$.} 
\end {figure}
\begin {figure}[!b]
\includegraphics{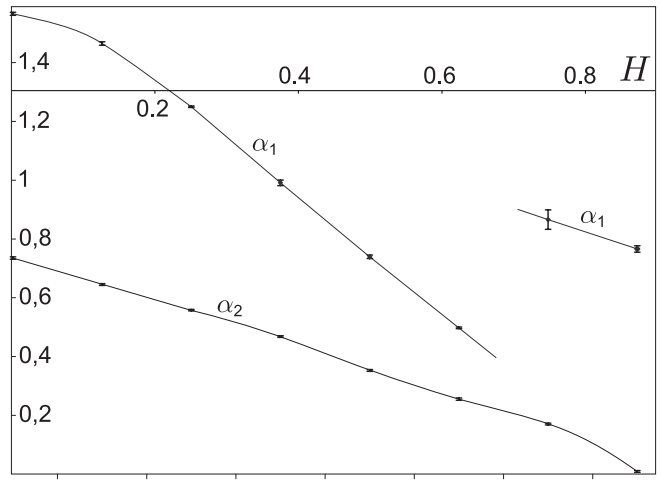}
\caption{Second and third largest exponents ($\alpha_1$ and $\alpha_2$) of the asymptotic expansion of the  
``oceanic coastline'' length are plotted versus 
the Hurst exponent $H$.} 
\end {figure}

Note that for $H \le -3/4$ (uncorrelated percolation), there is analytic result $d_c =91/48 \approx 1.8958$ \cite{Kapitulnik}, 
which is indistinguishable
(at our numerical uncertainty) from our result for $H=0$. 
While the author is not able to give a rigorous proof, it seems natural to assume that 
the function $d_f$ is a monotonically decreasing function of $H$. Indeed, a larger $H$ means a  stronger influence of 
long scales, which results in less twisted isolines, and a  lesser number of small islands (note that these arguments do not apply 
for $d_u$: excessive twists of the hull can efficiently 
screen and smooth the perimeter). This leads us to the conjecture $d_c \equiv 91/48$ for $H \le 0$.

In conclusion, we have introduced the concept of ``oceanic coastlines'', and shown that they are the 
extension of the concept of percolation clusters. The finding that their fractal dimension $d_c$ is clearly distinct from both the
dimension of single isolines $d_h$ and the dimension of the full set of isolines at a given height $d_f$ is particularly 
important, because in the context of Earth's landscapes, the coastlines of "oceanic islands" have been often 
mistakenly treated as the full set of isolines \cite{Mandelbrot}. Based on the numerical results, we have
conjectured that for $H \le 0$, the fractal dimension of percolation clusters is super-universal, $d_c \equiv 91/48$.
Last but not least, a novel and universal approach to the problem of interpreting the finite-size simulation results 
has been devised.

\end{document}